# Spectral decomposition of the Mueller matrix: A geometrical model

Colin J. R. Sheppard
Nanoscopy Group, Department of Nanophysics, Istituto Italiano di Tecnologia, Genova, 16163, Italy
Corresponding author: colinjrsheppard@gmail.com

**Abstract**
An arbitrary Müller matrix can be decomposed into a sum of up to four deterministic Müller-Jones matrices, with strengths given by the eigenvalues of an associated Hermitian matrix. A geometrical representation of the eigenvalues in terms of the matrix invariants is presented.

PACS: 4225Ja @ Polarization; 0760Fs @ Polarimeters and ellipsometers; 7820Fm @ Birefringence

The Müller matrix is the most common way to specify the complete description of the transformation of polarization state by an optical system or material. The Müller matrix is a $4 \times 4$ matrix of 16 real elements. An important approach to recognizing the significance of its structure is to separate it into the sum of up to four deterministic Müller-Jones matrices, a process called the spectral decomposition [1, 2]. This is a parallel decomposition, in contrast to the Lu-Chipman series decomposition [3]. The spectral decomposition is achieved by finding the eigenvalues of an associated Hermitian matrix. The eigenvalues are real and non-negative for a physically realizable matrix. So a negative eigenvalue is evidence of experimental error or noise. There are two different Hermitian matrices that have been reported in the literature [4], one in the Pauli representation like the Stokes parameters and the Müller matrix [5], and one in the standard Cartesian representation like the polarization coherency matrix (analogous to the density matrix in quantum mechanics) and the Jones matrix [6, 7]. We denote these two Hermitian matrices $\mathbf{G}$ and $\mathbf{H}$, respectively. The square roots of the eigenvalues also give the strength of four components, each described by a deterministic Jones matrix (or its equivalent in the Pauli representation, sometimes called the C-matrix [8]), which can be summed incoherently to describe a non-deterministic system.

We require to calculate the real eigenvalues of a $4 \times 4$ Hermitian matrix. First we briefly consider cases of lower order. The eigenvalues of a $2 \times 2$ Hermitian matrix can be obtained simply in an analytical form for the general case, by solving the quadratic characteristic equation. A $3 \times 3$ Hermitian matrix arises as the three-dimensional (3D) generalization of the polarization coherency matrix. For the $3 \times 3$ case, the usual solution of the corresponding cubic characteristic equation is not the most appropriate for real eigenvalues, as it is in the form of the sum of two cube roots, each with an imaginary part that cancels out in the summation. An alternative solution for real roots is a trigonometric solution, equivalent to a model in which the magnitudes of the three eigenvalues are given by the projection on to the $x$-axis of the vertices of an equilateral triangle [9-11]. The position of the center along the $x$-axis, the radius of the triangle's circumscribing circle, and the angle of rotation of the triangle, are all determined by the invariants of the Hermitian matrix. This model is equivalent to representing the eigenvalues in a ternary triangle plot, where the radius and angle of rotation are polar coordinates [11].

For the $4\times 4$ case, various different forms of analytical solution exist. But one model is a generalization of that described for the $3\times 3$ case. The values of the four eigenvalues are given by projections of the vertices of a regular tetrahedron [12-14]. This solution is then equivalent to representing the eigenvalues in a 3D tetrahedral quaternary plot, the normalized eigenvalues being barycentric coordinates.

The auxiliary equation of the matrix $\mathbf{A}$, which can be taken as either $\mathbf{G}$ or $\mathbf{H}$, is in general

$$x^4 + bx^3 + cx^2 + dx + e = 0, \tag{1}$$

where

$$\begin{aligned}
b &= -\mathrm{tr}\mathbf{A}, \\
c &= \tfrac{1}{2}\left[(\mathrm{tr}\mathbf{A})^2 - \mathrm{tr}\mathbf{A}^2\right], \\
d &= \tfrac{1}{6}\left[3\,\mathrm{tr}\mathbf{A}\,\mathrm{tr}\mathbf{A}^2 - (\mathrm{tr}\mathbf{A})^3 - 2\,\mathrm{tr}\mathbf{A}^3\right], \\
e &= \det \mathbf{A} = \Delta,
\end{aligned} \tag{2}$$

are expressed in terms of four matrix invariants, invariant under unitary transformations. They are therefore the same for both $\mathbf{G}$ and $\mathbf{H}$. The eigenvalues $\lambda_\alpha, 0 \le \alpha \le 3$ satisfy the conditions

$$\begin{aligned}
\lambda_0 + \lambda_1 + \lambda_2 + \lambda_3 &= \mathrm{tr}\mathbf{A} = 2M_{00}, \\
\lambda_0^2 + \lambda_1^2 + \lambda_2^2 + \lambda_3^2 &= \mathrm{tr}\mathbf{A}^2 \\
\lambda_0^3 + \lambda_1^3 + \lambda_2^3 + \lambda_3^3 &= \mathrm{tr}\mathbf{A}^3 \\
\lambda_0^4 + \lambda_1^4 + \lambda_2^4 + \lambda_3^4 &= \mathrm{tr}\mathbf{A}^4 \\
\lambda_0 \lambda_1 \lambda_2 \lambda_3 &= \Delta,
\end{aligned} \tag{3}$$

where $M_{00}$ is the first element of the Müller matrix. The degree of polarimetric purity is [2]

$$P^2 = \frac{1}{3}\left[\frac{4\,\mathrm{tr}\mathbf{A}^2}{(\mathrm{tr}\mathbf{A})^2} - 1\right], 0 \le P \le 1, \tag{4}$$

the two limiting cases being an ideal depolarizer, corresponding to $P=0$, $\mathrm{tr}\mathbf{A}^2 = \tfrac{1}{4}(\mathrm{tr}\mathbf{A})^2$, and a deterministic system corresponding to $P=1$, $\mathrm{tr}\mathbf{A}^2 = (\mathrm{tr}\mathbf{A})^2$. In analogy with the treatment for 3D polarization coherency, we can also introduce another two purity measures,

$$Q^3 = \frac{1}{15}\left[\frac{16\,\mathrm{tr}\mathbf{A}^3}{(\mathrm{tr}\mathbf{A})^3} - 1\right], 0 \le Q \le 1, \tag{5}$$

[11] and, in analogy to Barakat's measure for degree of polarization [15],

$$B^2 = 1 - \frac{256\Delta}{(\mathrm{tr}\mathbf{A})^4}, 0 \le B \le 1. \tag{6}$$

The three measures, $P, Q, B$ are sufficient to determine the state of purity of $\mathbf{H}$ and $\mathbf{G}$, and therefore $\mathbf{M}$. Alternatively, we can use the three measures $P, Q, S$, where $S$ is given by

$$S^4 = \frac{1}{63}\left[\frac{64\,\mathrm{tr}\mathbf{A}^4}{(\mathrm{tr}\mathbf{A})^4} - 1\right], 0 \leq S \leq 1, \tag{7}$$

where

$$\mathrm{tr}\mathbf{A}^4 = \tfrac{1}{6}\left[(\mathrm{tr}\mathbf{A})^4 - 6(\mathrm{tr}\mathbf{A})^2\,\mathrm{tr}\mathbf{A}^2 + 8\,\mathrm{tr}\mathbf{A}\,\mathrm{tr}\mathbf{A}^3 + 3(\mathrm{tr}\mathbf{A}^2)^2 - 24\Delta\right], \tag{8}$$

and

$$B^2 = 36P^2 - 18P^4 - 80Q^3 + 63S^4. \tag{9}$$

A quartic equation is usually solved by first constructing the reduced quartic equation by substituting $x = y - b/4$, to give [14]

$$y^4 + py^2 + qy + r = 0, \tag{10}$$

where

$$\begin{aligned} p &= \tfrac{1}{8}(8c - 3b^2), \\ q &= \tfrac{1}{8}(b^3 - 4bc + 8d), \\ r &= \tfrac{1}{256}(-3b^4 + 256e - 64bd + 16b^2c). \end{aligned} \tag{11}$$

From this reduced quartic equation we can generate the so-called resolvent cubic [14]:

$$y^3 + \tfrac{1}{2}py^2 + \tfrac{1}{16}(p^2 - 4r)y - \tfrac{1}{64}q^2 = 0, \tag{12}$$

with roots $g_1^2, g_2^2, g_3^2$, $g_1 \geq g_2 \geq g_3$, where

$$g_1 g_2 g_3 = -\tfrac{1}{8}q = \tfrac{1}{128}\mathrm{tr}\mathbf{A}^3(5Q^3 - 3P^2). \tag{13}$$

The eigenvalues of the original quartic are then

$$\begin{aligned} \lambda_0 &= \tfrac{1}{4}\mathrm{tr}\mathbf{A} + g_1 + g_2 + g_3, \\ \lambda_1 &= \tfrac{1}{4}\mathrm{tr}\mathbf{A} + g_1 - g_2 - g_3, \\ \lambda_2 &= \tfrac{1}{4}\mathrm{tr}\mathbf{A} + g_2 - g_3 - g_1, \\ \lambda_3 &= \tfrac{1}{4}\mathrm{tr}\mathbf{A} + g_3 - g_1 - g_2. \end{aligned} \tag{14}$$

In order for the condition $\lambda_0 \geq \lambda_1 \geq \lambda_2 \geq \lambda_3 \geq 0$ to be satisfied, $g_1, g_2$ must both be positive, so the sign of $g_3$ is determined by Eq. 12.

It has been shown that the eigenvalues are given by the projection on to the $x$-axis of the four vertices of a regular tetrahedron [14], with circumscribing radius

$$R = \sqrt{-\frac{3p}{2}} = \tfrac{3}{4}\sqrt{\frac{3b^2 - 8c}{3}} = \tfrac{3}{4}P\,\mathrm{tr}\mathbf{A}, \tag{15}$$

side $s = \sqrt{3/2}\,P\,\mathrm{tr}\mathbf{A}$, and center of gravity at $y = 0$, as shown in Fig. 1, and can therefore be written

$$\lambda_\alpha = \tfrac{1}{4}\mathrm{tr}\mathbf{A}(1 + 3P\cos\theta_\alpha), \tag{16}$$

where $\theta_\alpha$ is the angle between the line from the center of the tetrahedron to the vertex and the $x$-axis. Then $\pm g_j$ are given by the projections of the mid-points of the sides of the tetrahedron, which together form a regular octahedron, on to the $x$-axis.

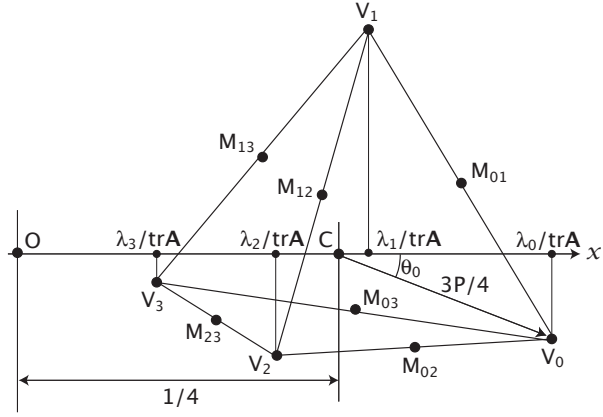

Fig. 1. The eigenvalues of **G** or **H** are given by projecting from the four vertices $V_\alpha$ of a regular tetrahedron on to the $x$-axis. The center $C$ of the tetrahedron is displaced a normalized distance of $\frac{1}{4}$ along the axis. The mid-points $M_{\alpha\beta}$ of the edges of the tetrahedron form a regular octahedron, and their projections give the values of $g_j$.

We can distinguish four special cases, as shown in Table 1, one with four equal non-zero eigenvalues, and the others with 3, 2, or 1 zero-valued eigenvalues. Case A corresponds to the deterministic case. Case D corresponds to an ideal depolarizer. Gil introduced three indices of purity, $P_1, P_2, P_3$, values for which are also given [2].

**Table 1. Four special cases for the eigenvalues.**

|   | A | B | C | D |
|---|---|---|---|---|
| $l$/tr**A** | (1, 0, 0, 0) | 1/2 (1,1,0,0) | 1/3 (1,1,1,0) | 1/4 (1,1,1,1) |
| P | 1 | 0.577 | 0.333 | 0 |
| Q | 1 | 0.585 | 0.373 | 0 |
| S | 1 | 0.699 | 0.384 | 0 |
| B | 1 | 1 | 1 | 0 |
| $P_1$ | 1 | 0 | 0 | 0 |
| $P_2$ | 1 | 1 | 1/3 | 0 |
| $P_3$ | 0 | 0 | 1/3 | 0 |

Gil's indices of purity are defined as [2]
$$P_1 = \frac{\lambda_0 - \lambda_1}{\text{tr}\mathbf{A}}, P_2 = \frac{(\lambda_0 + \lambda_1) - (\lambda_2 + \lambda_3)}{\text{tr}\mathbf{A}}, P_3 = \frac{\lambda_2 - \lambda_3}{\text{tr}\mathbf{A}}, \qquad (17)$$
giving
$$\begin{aligned}\lambda_0 &= \tfrac{1}{4}\text{tr}\mathbf{A}(1 + P_2 + 2P_1),\\ \lambda_1 &= \tfrac{1}{4}\text{tr}\mathbf{A}(1 + P_2 - 2P_1),\\ \lambda_2 &= \tfrac{1}{4}\text{tr}\mathbf{A}(1 - P_2 + 2P_3),\\ \lambda_3 &= \tfrac{1}{4}\text{tr}\mathbf{A}(1 - P_2 - 2P_3),\end{aligned} \qquad (18)$$
so that they are related to the $g_j$'s:
$$g_1 = \tfrac{1}{4}\text{tr}\mathbf{A}\, P_2, g_2 = \tfrac{1}{4}\text{tr}\mathbf{A}(P_1 + P_3), g_3 = \tfrac{1}{4}\text{tr}\mathbf{A}(P_1 - P_3). \qquad (19)$$

We have the relationship between Gil's indices and the degree of polarimetric purity [2],

$$P^2 = \tfrac{1}{3}(2P_1^2 + P_2^2 + 2P_3^2). \tag{20}$$

The eigenvalues can be plotted on a tetrahedral plot, also called a barycentric or simplex plot. We introduce Cartesian coordinates $X,Y,Z$ in 3D space, as in Fig. 2,

$$X = \frac{\sqrt{3}}{4}P_2, Y = \frac{\sqrt{6}}{4}P_1, Z = \frac{\sqrt{6}}{4}P_3, \tag{21}$$

recognizing that $P_1, P_2, P_3$ are mutually orthogonal in this representation. Then

$$\left(\frac{\lambda_0}{\mathrm{tr}\mathbf{A}} - \tfrac{1}{4}\right) = \frac{1}{\sqrt{3}}X + \sqrt{\tfrac{2}{3}}Y, \quad \left(\frac{\lambda_1}{\mathrm{tr}\mathbf{A}} - \tfrac{1}{4}\right) = \frac{1}{\sqrt{3}}X - \sqrt{\tfrac{2}{3}}Y,$$
$$\left(\frac{\lambda_2}{\mathrm{tr}\mathbf{A}} - \tfrac{1}{4}\right) = -\frac{1}{\sqrt{3}}X + \sqrt{\tfrac{2}{3}}Z, \left(\frac{\lambda_3}{\mathrm{tr}\mathbf{A}} - \tfrac{1}{4}\right) = -\frac{1}{\sqrt{3}}X - \sqrt{\tfrac{2}{3}}Z, \tag{22}$$

so that

$$\sqrt{X^2 + Y^2 + Z^2} = \tfrac{3}{4}P \tag{23}$$

represents a spherical radius. Cases A, B and C are shown in Fig. 2. Case C corresponds to a point out of the plane, at $Z = 1/(2\sqrt{6})$. Points for $\lambda_0 \geq \lambda_1 \geq \lambda_2 \geq \lambda_3 \geq 0$ are contained in a tetrahedral (irregular) volume ABCD, with face ABD shown shaded. There are 24 such regions in the complete tetrahedron that applies for any values of normalized eigenvalues. Gil describes the conditions on the values of the indices of purity in detail [2].

Thus we see that the solution for the eigenvalues based on the projection from the vertices of a regular tetrahedron leads directly to the representation of the complete polarization transformation state in a tetrahedral plot.

Comparing the solutions in Eqs. 15 and 21, we can obtain the conditions

$$\cos\theta_0 + \cos\theta_1 + \cos\theta_2 + \cos\theta_3 = 0,$$
$$\cos^2\theta_0 + \cos^2\theta_1 + \cos^2\theta_2 + \cos^2\theta_3 = \tfrac{4}{3}. \tag{24}$$

The values $\cos\theta_\alpha$ are like generalized direction cosines. We can also specify a particular polarization transformation condition by spherical polar coordinates, $\tfrac{3}{4}P, \chi, \phi$:

$$X = \tfrac{3}{4}P\cos\chi,$$
$$Y = \tfrac{3}{4}P\sin\chi\cos\phi, \tag{25}$$
$$Z = \tfrac{3}{4}P\sin\chi\sin\phi,$$

and can then obtain various relationships between the coordinate systems, such as

$$\cos\theta_0 = \frac{1}{\sqrt{3}}\cos\chi + \sqrt{\tfrac{2}{3}}\sin\chi\cos\phi,$$
$$\cos\theta_1 = \frac{1}{\sqrt{3}}\cos\chi - \sqrt{\tfrac{2}{3}}\sin\chi\cos\phi,$$
$$\cos\theta_2 = -\frac{1}{\sqrt{3}}\cos\chi + \sqrt{\tfrac{2}{3}}\sin\chi\sin\phi, \tag{26}$$
$$\cos\theta_3 = -\frac{1}{\sqrt{3}}\cos\chi - \sqrt{\tfrac{2}{3}}\sin\chi\sin\phi.$$

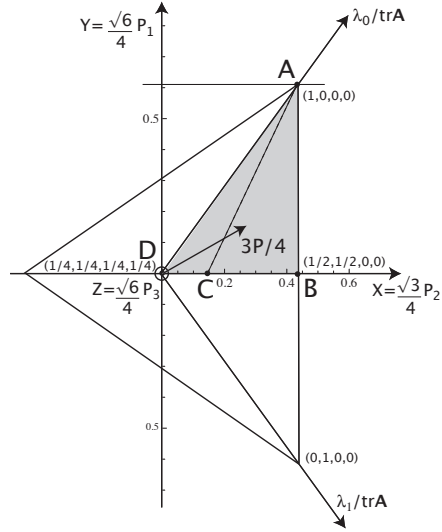

Fig. 2. An XY section through a 3D plot showing the polarization transformation condition at a point on a sphere radius $\tfrac{3}{4}P$ as a quaternary plot. Cases A, B and D of Table 1 are in the plane. Case C is above the plane, forming an irregular tetrahedron ABCD.

Next we consider the values of the eigenvalues in terms of the matrix invariants. First, from Eqs. 2, 11, 13, 19 and 21, we obtain the relationships

$$P_2(P_1^2 - P_3^2) = \frac{\left[(\mathrm{tr}\mathbf{A})^3 - 6\,\mathrm{tr}\mathbf{A}\,\mathrm{tr}\mathbf{A}^2 + 8\,\mathrm{tr}\mathbf{A}^3\right]}{3(\mathrm{tr}\mathbf{A})^3} = \tfrac{1}{2}(5Q^3 - 3P^2)$$
$$= \frac{32\sqrt{3}}{9} X(Y^2 - Z^2) = \frac{64 g_1 g_2 g_3}{\mathrm{tr}\mathbf{A}}. \quad (27)$$

Eqs. 20, 27 then give an expression for $Q^3$ in terms of the indices of purity,

$$5Q^3 = 2P_1^2(1 + P_2) + P_2^2 + 2P_3^2(1 - P_2). \quad (28)$$

The roots $g_1^2, g_2^2, g_3^2$ can be calculated from the resolvent cubic. As they are necessarily real, the trigonometric method is preferable [10, 11]. We find that

$$g_j^2 = \tfrac{1}{16}(\mathrm{tr}\mathbf{A})^2 \left\{ P^2 + \left(20Q^3 - 6P^2 + 7P^4 - 21S^4\right)^{1/2} \cos\left[\psi + \frac{2\pi(j-1)}{3}\right] \right\}, \quad (29)$$

where

$$\psi = \frac{1}{3}\arccos\left[\frac{30P^2 Q^3 - 9P^4 + 17P^6 + 25Q^6 - 63P^2 S^4}{\left(20Q^3 - 6P^2 + 7P^4 - 21S^4\right)^{3/2}}\right], \quad (30)$$

or, alternatively,

$$g_j^2 = \tfrac{1}{16}(\mathrm{tr}\mathbf{A})^2 \left\{ P^2 + \frac{1}{\sqrt{3}}\left(18P^2 + 3P^4 - 20Q^3 - B^2\right)^{1/2} \cos\left[\psi + \frac{2\pi(j-1)}{3}\right] \right\}, \quad (31)$$

where

$$\psi = \frac{1}{3}\arccos\left[\frac{3\sqrt{3}\left(27P^4 - P^6 - 50P^2 Q^3 + 25Q^6 - P^2 B^2\right)}{\left(18P^2 + 3P^4 - 20Q^3 - B^2\right)^{3/2}}\right]. \quad (32)$$

The eigenvalues can then be calculated from Eq. 14, or the indices of purity from Eq. 17. We are not suggesting that this is an efficient way to calculate the eigenvalues numerically for a particular case, however. The value of the approach is that it leads to an intuitive appreciation of the behavior, and provides a connection with both the degree of polarimetric purity and the indices of purity. On the other hand, the usual general solution is too complicated to appreciate the structure and trends.

Next we describe some special cases. The case of a single zero eigenvalue, $\lambda_3 = 0$, corresponds to the condition $P_2 + 2P_3 = 1$, defining the plane ABC, corresponding to $B = 1$. In this case the quartic equation reduces to a cubic equation. The plane ACD with equation $P_2 = P_1 + P_3$ corresponds to $g_1 = g_2$, giving $\lambda_1 = \lambda_2$. The plane BCD is $P_1 = 0$, giving $\lambda_0 = \lambda_1$, and $g_2 + g_3 = 0$. The plane ABD is $P_3 = 0$, giving $\lambda_2 = \lambda_3$, and $g_2 = g_3$. The case of two zero eigenvalues, $\lambda_2 = \lambda_3 = 0$, corresponds to the line $P_2 = 1, P_3 = 0$, with $g_2 = g_3$, and $g_1 = \tfrac{1}{4}\mathrm{tr}\mathbf{A}$. Then $3(P^2 - 1) = 2(P_1^2 - 1)$. In this case the quartic equation reduces to a quadratic.

For completeness, we describe briefly here the generation of the matrices $\mathbf{G}$ and $\mathbf{H}$. These are defined as expansions of the Müller matrix, with elements $M_{\mu\nu}$, as [4-6]

$$\mathbf{G} = \sum_{\mu,\nu} M_{\mu\nu} \mathbf{\Gamma}_{\mu\nu},$$
$$\mathbf{H} = \sum_{\mu,\nu} M_{\mu\nu} \mathbf{\Psi}_{\mu\nu}, \tag{33}$$

where the basis functions are

$$\mathbf{\Gamma}_{\mu\nu} = \mathbf{\Lambda}^\dagger (\boldsymbol{\sigma}_\mu \otimes \boldsymbol{\sigma}_\nu^*) \mathbf{\Lambda},$$
$$\mathbf{\Psi}_{\mu\nu} = (\boldsymbol{\sigma}_\mu \otimes \boldsymbol{\sigma}_\nu^*), \tag{34}$$

the asterisk denoting complex conjugate, and the dagger denoting conjugate transpose. Here the Pauli spin matrices are taken in the optical order:

$$\boldsymbol{\sigma}_\mu = \frac{1}{\sqrt{2}} \left[ \begin{pmatrix} 1 & 0 \\ 0 & 1 \end{pmatrix}, \begin{pmatrix} 1 & 0 \\ 0 & -1 \end{pmatrix}, \begin{pmatrix} 0 & 1 \\ 1 & 0 \end{pmatrix}, \begin{pmatrix} 0 & -i \\ i & 0 \end{pmatrix} \right], \tag{35}$$

and the matrix $\mathbf{\Lambda}$ is formed from the column vectors of the Pauli matrices [7],

$$\mathbf{\Lambda} = \begin{pmatrix} \hat{\boldsymbol{\sigma}}_0 & \hat{\boldsymbol{\sigma}}_1 & \hat{\boldsymbol{\sigma}}_2 & \hat{\boldsymbol{\sigma}}_3 \end{pmatrix} = \frac{1}{\sqrt{2}} \begin{pmatrix} 1 & 1 & 0 & 0 \\ 0 & 0 & 1 & -i \\ 0 & 0 & 1 & i \\ 1 & -1 & 0 & 0 \end{pmatrix}. \tag{36}$$

Then

$$\mathbf{G} = \mathbf{U}\mathbf{D}\mathbf{U}^\dagger = \sum_{\alpha=0}^{3} \lambda_\alpha \overline{\mathbf{G}}_\alpha, \tag{37}$$

where $\mathbf{U}$ is the matrix of eigenvectors, $\mathbf{D}$ is the diagonal matrix formed from the eigenvalues, and the normalized, deterministic matrices $\overline{\mathbf{G}}_\alpha$ are formed from the eigenvectors:

$$\overline{\mathbf{G}}_\alpha = \mathbf{u}_\alpha \mathbf{u}_\alpha^\dagger = \mathbf{u}_\alpha \otimes \mathbf{u}_\alpha^*. \tag{38}$$

Similarly

$$\mathbf{H} = \mathbf{V}\mathbf{D}\mathbf{V}^\dagger = \sum_{\alpha=0}^{3} \lambda_\alpha \overline{\mathbf{H}}_\alpha, \quad \overline{\mathbf{H}}_\alpha = \mathbf{v}_\alpha \mathbf{v}_\alpha^\dagger = \mathbf{v}_\alpha \otimes \mathbf{v}_\alpha^*. \tag{39}$$

The eigenvalues of $\mathbf{G}$ give the strengths of four normalized, fully determinstic Müller-Jones components $\overline{\mathbf{M}}_\alpha$ of the Müller matrix:

$$\mathbf{M} = \sum_{\alpha=0}^{3} \lambda_\alpha \bar{\mathbf{M}}_\alpha,$$

$$\bar{\mathbf{M}}_\alpha = \sum_{\mu,\nu} \left[\bar{\mathbf{G}}_\alpha\right]_{\mu\nu} \mathbf{\Gamma}_{\mu\nu}, \qquad (40)$$

A similar approach can be based on the eigenvectors of $\mathbf{H}$, where

$$\bar{\mathbf{M}}_\alpha = \sum_{\mu,\nu} \left[\bar{\mathbf{H}}_\alpha\right]_{\mu\nu} \mathbf{\Psi}_{\mu\nu}. \qquad (41)$$

Finally, we mention that the geometrical model given here applies equally to other applications of $4 \times 4$ Hermitian matrices, which have real, non-negative eigenvalues, in quantum optics for example.